\documentclass[twocolumn, prl]{revtex4}

\usepackage{graphicx}
\usepackage{dcolumn}
\usepackage{bm}

\begin{document}

\title{ The Plaquette Ground State of the Shastry-Sutherland Model}

\author {S. Moukouri }

\affiliation{ Department of Physics and  Michigan Center for 
          Theoretical Physics \\
         University of Michigan 2477 Randall Laboratory, Ann Arbor MI 48109}

\begin{abstract}

I use the two-step density-matrix renormalization group method based on
two-leg ladder expansion to show numerical evidence of a plaquette ground 
state for $J_2=1.3J_1$ in the Shastry-Sutherland model. I argue that the 
DMRG method is  very efficient in the strong frustration regime of 
two-dimensional spin models where a spin-Peierls ground state is expected 
to occur. It is thus complementary to quantum Monte Carlo algorithms, which 
are known to work well in the small frustration regime but which are plagued
by the sign problem in the strong frustration regime.
 
\end{abstract}

\maketitle

\begin{figure}
\begin{center}
$\begin{array}{c@{\hspace{0.5in}}c}
         \multicolumn{1}{l} {}\\ [-0.23cm]
\includegraphics[width=1.5 in, height=1.5 in]{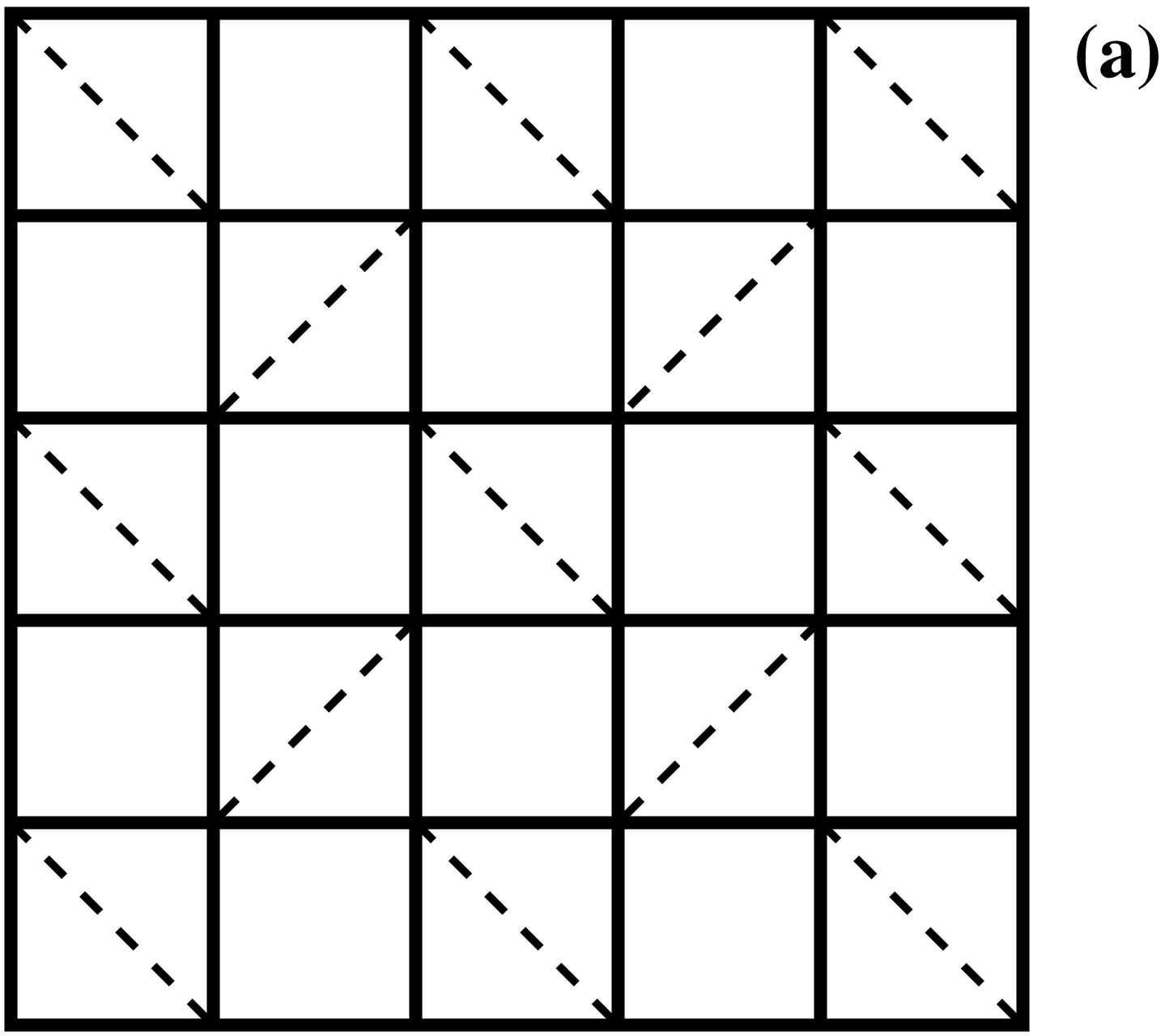}
\hspace{.75cm}
\includegraphics[width=0.5 in, height=1.5 in]{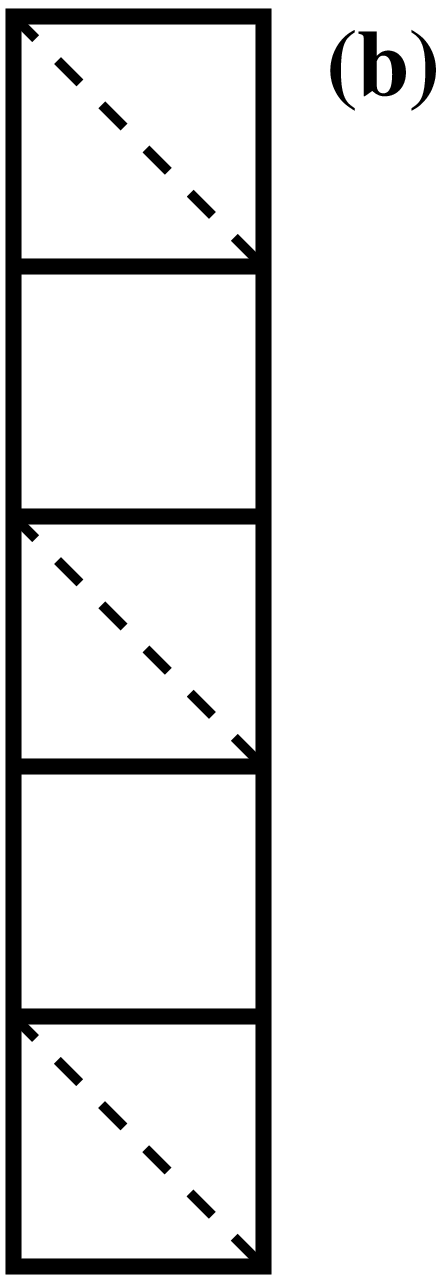}
\hspace{.75cm}
\includegraphics[width=0.75 in, height=1.5 in]{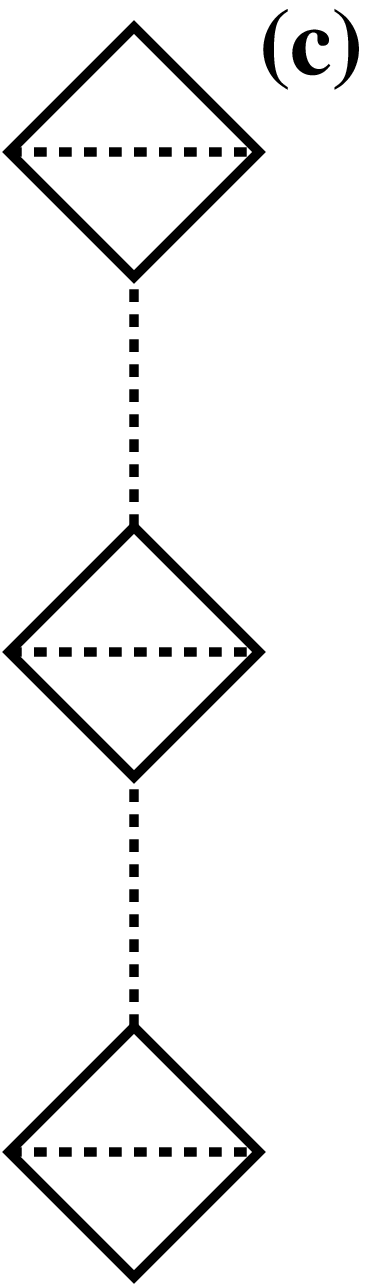}
\end{array}$
\end{center}
\caption{(a) Bond patterns in the Shastry-Sutherland lattice. The $J_1$ bonds
are between the nearest neighbors. The $J_2$ bonds are along every other
diagonal. (b) the two-leg ladder: starting point of the two-step DMRG approach
to the two-dimensional lattice.(c) Orthogonal dimers: starting point more 
adapted to the dimerized phase. It is obtained by covering the two-dimensional
lattice allong the diagonals.}
\vspace{0.5cm}
\label{lattice}
\end{figure}

A number of studies have been devoted to the Shastry-Sutherland model (SSM)
\cite{shastry,mila,miyahara,weihong,muller-hartmann,koga,marston,oitmaa,carpentier,lauchli,chung,liu}. This interest is motivated by the relevance of 
the SSM to the physics of the two-dimensional spin gap system 
$SrCu_2(BO_3)_2$\cite{kageyama}. The
SSM is a frustrated antiferromagnetic model on a square lattice whose
Hamiltonian is written as

\begin{eqnarray}
  H=J_1 \sum_{<i,j>}{\bf S}_i{\bf S}_j+
J_2 \sum_{[i,j]}{\bf S}_i{\bf S}_j,
\label{hamiltonian}
\end{eqnarray}

\noindent where $<i,j>$ represents nearest-neighbor sites and $[i,j]$
stands for the next-nearest neighbors on every other plaquette in the pattern
shown in Fig.\ref{lattice}. There is a general agreement that in the
weak frustration regime $J_1 \gg J_2$, the SSM is N\'eel ordered while
in the strong frustration regime $J_1 \ll J_2$, the model is a valence
bond solid. In fact, Shastry and Sutherland showed that for $J_2 > 2J_1$
the wavefunction made of the product of local orthogonal dimers is an exact 
eigenstate.
Numerical simulations based on series expansions and exact diagonalization
have pushed the dimer phase boundary down to $J_2 \approx 1.5 J_1$. The
estimated boundary of the N\'eel phase is $J_2 \approx 1.2 J_1$.

However, the nature of the ground state for $1.2 J_1 \alt J_2 \alt 1.5 J_1$ 
so far has remained controversial. A mean-field Schwinger boson approach 
found that the intermediate phase is helical \cite{mila}. 
Perturbation theory \cite{miyahara} and series expansion \cite{weihong}
studies predicted a 
direct first order transition between the N\'eel and the dimer phases. 
A large N study \cite{marston} also predicted a helical phase. In addition 
it suggested that a broader phase diagram contains a plaquette phase which 
might occur if fluctuactions were included. A subsequent series expansion 
analysis \cite{koga} predicted a plaquette phase with a spin gap. 
This conclusion was criticized in another series expansion study which 
suggested a possible
gapless phase whose nature was unclear \cite{oitmaa}. The existence 
of the intermediate phase 
was also suggested in a renormalization group analysis \cite{carpentier}.
 An exact diagonalization study on a $N=32$ site system has concluded to 
a plaquette phase. The Monte Carlo method, which is very effective for spin
systems in absence of frustration \cite{young}, is plagued by the sign problem 
in this regime of strong frustration. Knowledge of the exact ground
state phases is essential; this could serve as a 
starting point in variational investigations of the nature of 
superconductivity that might arise upon doping \cite{liu}. 

In this letter I present numerical evidence of the plaquette
phase at $J_2=1.3 J_1$. For this purpose, I will use the two-step
 density-matrix renormalization group (DMRG) method 
\cite{moukouri-TSDMRG,moukouri-TSDMRG2}. The DMRG \cite{white} has
provided a breakthrough in the study of quantum Hamiltonians in one dimension.
 Extensions of the DMRG to two-dimensional Hamiltonians have been less
effective. Liang and Pang \cite{pang} found that as the linear dimensions of the
system grow, the number of the reduced density matrix states needed to 
maintain accuracy grows exponentially. This problem is particularly severe for
 quantum antiferromagnets in their ordered phase. The spontaneous symmetry
breaking which takes place in the thermodynamic limit is due to the
collapse of an infinite number of excited states onto the ground state.
The implication for finite systems, in the parameter regime where long-range
order occurs, is a near degeneracy of a large number of low-lying states
with the ground state. Each state within this large set  would carry the 
same weight in the reduced density matrix. For this reason, standard DMRG 
simulations of spin Hamiltonians are limited to systems of about ten sites wide.

Recent developments by the author 
\cite{moukouri-TSDMRG,moukouri-TSDMRG2,checkerboard} 
have shown that the DMRG could be very useful for 2D models in the 
region of the parameter space where this technical difficulty is less 
severe or even absent. This occurs for instance for spatially anisotropic 
antiferromagnets or in the highly 
frustrated regime of isotropic magnets. In this latter case, general 
arguments from the large N approach \cite{read-sachdev} suggest that the ground 
state is a collection of weakly-coupled dimers or plaquettes. Presumably 
deep in the disordered phase, because of the presence of a spin gap, the 
ground state would be dominant in the reduced density matrix. 
This is more favorable to a DMRG simulation. It is usually in this regime 
that the sign problem is most severe in QMC simulations. 
Hence the DMRG would be complementary
to QMC for frustrated models with disordered phases with broken translational
symmetry. The excited states would become more and more important as the 
coupling is moved toward the boundary with the magnetically 
ordered phase. An approach based of these ideas recently has been applied 
to find the ground-state phase diagram of the checkerboard model 
\cite{checkerboard}. The same approach is applied here to the SSM.

\begin{figure}
\includegraphics[width=3. in, height=2. in]{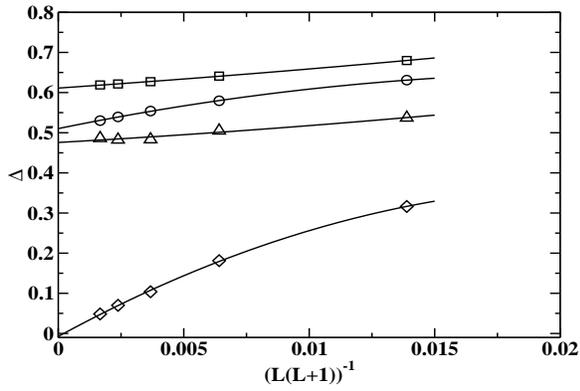}
\caption{ Gaps as a function of $L$: for a single ladder for $J_2=0$ (circles),
$J_2=1.3 J_1$ (squares), and for two-dimensional systems for $J_2=0$ (diamonds),
$J_2=1.3 J_1$ (triangles).}
\vspace{0.5cm}
\label{gaps}
\end{figure}

\begin{figure}
\includegraphics[width=3. in, height=2. in]{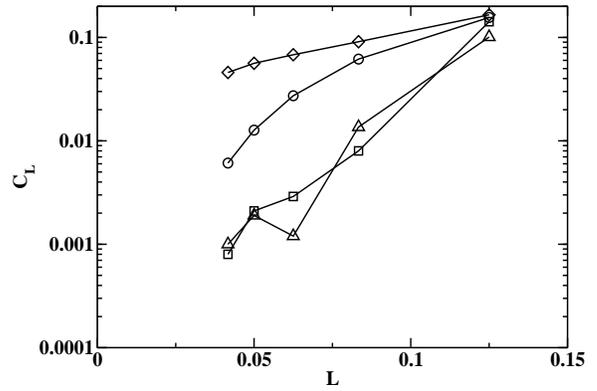}
\caption{Edge-to-center correlation as a function of $L_y$: for a single 
ladder for $J_2=0$ (circles), $J_2=1.3 J_1$ (squares), and for two-dimensional 
systems for $J_2=0$ (diamonds), $J_2=1.3 J_1$ (triangles).}
\vspace{0.5cm}
\label{corl}
\end{figure}

I start with a single two-leg ladder with $L$ rungs as shown in 
Fig.\ref{lattice}. By doing so, I
implicitly assume that the inter-ladder interactions are small. 
Strictly, this is true only in the magnetically disordered phase and 
will be justified a posteriori. However, such a starting point could 
also be justified qualitatively even for the magnetically ordered phase 
where inter-ladder interactions are not small. In the valence bond 
represensation of quantum antifferomagnets,
the wave function is written as

\begin{eqnarray} 
\Psi=\sum_{\alpha} c_{\alpha} \prod_{(ij)\in \{\alpha\}}|{\bf(ij)}\rangle,
\end{eqnarray}

\noindent where 
$|{\bf (ij)}\rangle=(|\uparrow \downarrow \rangle_{ij}-|\downarrow \uparrow \rangle_{ij})/\sqrt{2}$ 
is a dimer wave function between the sites 
$i$ and $j$, and $\{\alpha \}$ is a configuration of dimers. Such valence
bond representations are qualitatively good for both the disordered and
ordered phases \cite{liang}. In the present approach, two-leg ladders
are the building blocks. The wave function is written as

\begin{eqnarray}  
{\tilde \Psi}=\sum_{ladders}{\tilde c}_{ladders}\prod_{ladders} \Phi_{ladder},
\end{eqnarray}

\noindent where $\Phi_{ladder}$ is an eigenfunction of a single ladder 
Hamiltonian. Given that the lowest $\Phi_{ladder}$ is dominated by a product
of dimers, ${\tilde \Psi}$ bears some similarity with $\Psi$.
However, the set of $\Phi_{ladder}$'s includes excited
states on the ladder, the structure of ${\tilde \Psi}$ is thus much
more complex than a simple short-range dimer expansion. 
When the ground state is made of weakly-coupled plaquettes
or dimers,  it would be expected that this representation would yield
quantitatively good results as well. But the essential point is that this
approximation does not necessarily assume that the ground state is 
disordered. It will be shown below that a magnetically ordered state can be
reached as well, though with less accuracy than in the disordered phase.  

\begin{figure}
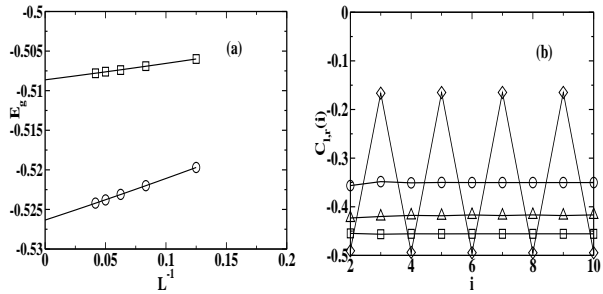

\begin{center}
$\begin{array}{c@{\hspace{0.5in}}c}
         \multicolumn{1}{l} {}\\ [-0.23cm]
\includegraphics[width=1.5 in, height=1.5 in]{eg-lad.eps}
\hspace{0.25cm}
\includegraphics[width=1.5 in, height=1.5 in]{corloc.eps}
\end{array}$
\end{center}
\caption{ (a) Ground state energy of two-leg ladders: for the Shastry-Sutherland
model at $J_2=1.3 J_1$ (circles) and for the checkerboard model at $J_2=J_1$ 
(squares). (b) Short-range correlations for the Shastry-Sutherland two-leg 
ladder with $L_x=24$: for $J_2=0$, $C_{l_{n,d}}$ (circles), $C_r$ (squares); for $J_2=1.3 J_1$,
$C_{l_{n,d}}$ (diamonds),  $C_r$ (triangles).}
\vspace{0.5cm}
\label{short}
\end{figure}

 The results for an isolated two-leg ladder were obtained for $J_2=0$ 
and for $J_2=1.3 J_1$. The conventional DMRG is known to yield highly
accurate results for the ground state and the low-lying states \cite{azzouz}.
A set of the low-lying $\Phi_{ladder}$ is obtained by targeting the
spin sectors from $S=0$ to $S=\pm4$ and by keeping up to $m=144$ states. This
is enough to maintain the truncation error below $10^{-6}$ in all cases. 
There is a spin gap in the thermodynamic limit as in two cases 
seen in Fig.\ref{gaps}. The center-to-end correlation functions 
$C_L=\langle {\bf S}_{L/2+1,r} {\bf S}_{L,r} \rangle$, where $r=1,2$ is 
the leg index, shown in Fig.\ref{corl} decay exponentially in both cases. 
But the short-range correlations in Fig.\ref{short} reveal that 
$J_2=0$ and $J_2=1.3 J_1$ belong to two different phases of the ladder. 
When $J_2=0$, the dominant short-range
correlation are the rung dimers $C_{r}(i)=\langle {\bf S}_{i,1}
{\bf S}_{i,2} \rangle$. $C_r$ is stronger than the correlations along
a leg $C_{l_{n,d}}(i)=\langle {\bf S}_{i,r}{\bf S}_{i+1,r} \rangle_{n,d}$ 
for plaquettes with no diagonal bond (n) or with a diagonal bond (d); 
when $J_2=0$, $C_{l_n}=C_{l_d}$. Both $C_r(i)$ and $C_{l_{n,d}}(i)$ are 
independent of $i$, except for small variations at the boundary. 
For $J_2=1.3 J_1$, $|C_{l_n}| > |C_{l_d}|$ and the bond pattern shows strong 
alternations as function of $i$ as seen in Fig.\ref{short}. At the same time, 
$C_r \approx C_{l_n}$ indicating that the system is now in the plaquette
phase. The plaquette-plaquette interaction is given by $|C_{l_d}|$.
It is not negligible as in the checkerboard ladder \cite{checkerboard}. 
It is about one third of the intra-plaquette interaction in the 
Shastry-Sutherland ladder, while it is only $5\%$ of the intra-plaquette 
in the checkerboard ladder at the isotropic point $J_2=J_1$.
This is also seen in the ground state energy $E_g$ of the ladders shown in
Fig.\ref{short}. In the checkerboard ladder, there is a relatively small
renormalization of $E_g=-0.5086 J_1$, which is not very far from 
$E_g=-0.5000 J_1$ of an isolated plaquette. The renormalization is more 
important in the Shastry-Sutherland ladder where $E_g=-0.5263 J_1$. 
 Nevertheless, the value of $C_{l_d}$ shows that even the SSM is in 
the weak-coupling regime of plaquette-plaquette interaction.

\begin{figure}
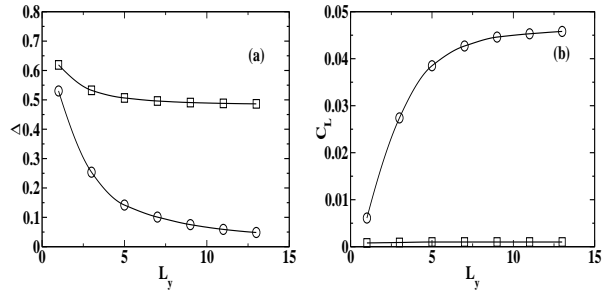

\begin{center}
$\begin{array}{c@{\hspace{0.5in}}c}
         \multicolumn{1}{l} {}\\ [-0.23cm]
\includegraphics[width=1.5 in, height=1.5 in]{rengap.eps}
\hspace{0.25cm}
\includegraphics[width=1.5 in, height=1.5 in]{rencorl.eps}
\end{array}$
\end{center}
\caption{ Irrelevant (squares, $J_2=1.3 J_1$) and relevant (circles, $J_2=0$) 
flows from the single ladder as function of $L_y$: (a) for the spin gap,
(b) for the center-to-edge correlation function.}
\vspace{0.5cm}
\label{ren}
\end{figure}

The second step of the two-step DMRG consists in projecting the 
Hamiltonian (\ref{hamiltonian}) onto the basis states of 
the tensor product of $\Phi_{ladder}$'s and solving the resulting
effective Hamiltonian, which is one-dimensional 
( in the transverse direction) with the usual DMRG. 
Strictly speaking, this can only rigorously be justified if
the inter-ladder coupling is small. Apparently this is not the case 
for Hamiltonian (\ref{hamiltonian}), since neither $J_1$ nor $J_2$ is
always small. But as seen above, effective small interactions can
be generated by the inherent competition between $J_1$ and $J_2$.
 The coupling between ladders in the SSM is also given by 
$|C_{l_d}|$, as can be seen in Fig.\ref{lattice}. 
In Ref\cite{moukouri-TSDMRG2},it has been shown that when the ratio,
$\rho=\delta E/J_{eff}$ between the bandwidth of the states kept, and the 
effective transverse coupling, which is $J_{eff}\approx |C_{l_d}|J_1$, 
is large ($\rho \agt 4$), the two-step DMRG yields results which 
are comparable to those of QMC. Typical values of $\rho$ for the largest 
systems studied are $5$ for $J_2=1.3 J_1$, and $2$ for $J_2=0$. This justifies 
the isolated ladder starting point for $J_2=1.3J_1$. 
I study lattices with $L_x\times L_y=L\times (L+1)=8\times9$ to
$24 \times 25$. Analysis of the performance of this approach have
been discussed in previous publications \cite{moukouri-TSDMRG,checkerboard}. 
When the inter-ladder is turned on, the flows as the function of the number
of ladders of $\Delta$ and $C_L$ for $J_2=0$ and $J_2=1.3 J_1$, shown 
in Fig.\ref{ren} for $L\times(L+1)=24 \times 25$, are very different.

For $J_2=0$, as expected from the existence of long-range order, 
$\Delta$ decays rapidly as $L_y$ increases. $\Delta=0$ in the thermodynamic
limit as seen in Fig.\ref{gaps}. At the same time, $C_L$ grows away
from its small value found in the two-leg ladder. $C_L$ becomes
finite in the thermodynamic limit. The extrapolated order parameter 
$M=\sqrt{C_{\infty}}$ is found to be $M=0.1738$. This is somewhat lower than
the QMC results \cite{sandvik} $M=0.3070$. Part of this discrepancy is due 
to the use of open boundary conditions which yield an undervaluated $C_L$.
 Better extrapolations can be obtained if the lattice sizes are reduced,  
 and if periodic boundary condition are used. 
This relatively poor performance of the DMRG deep in the magnetically 
ordered phase is a consequence of the fundamental limitations of the DMRG 
when faced with an exponentially dense low energy spectrum, as discussed in the 
introduction. Nevertheless, this results shows that the approximation by
${\tilde \Psi}$ of the exact wave function  retains the correct qualitative 
behavior. Hence the DMRG could still be very useful in the magnetic regime
as well. 

However, when $J_2=1.3 J_1$, the inter-ladder coupling does not qualitatively 
affect the physics of a single ladder which is itself that of nearly isolated
plaquettes as seen above. The irrelevant flows for $\Delta$ and
$C_L$ with the number of ladders are shown in Fig.\ref{ren}. 
$\Delta$ for the two-dimensional system is renormalized
by about $20 \%$ from its single ladder value.  The extrapolated gap 
$\Delta=0.4758 J_1$ is lower than $\Delta=0.67 J_1$ found in the 
checkerboard model at the isotropic point\cite{checkerboard}. 
This is consistent with the 
fact that inter-plaquette
interactions are more important in the SSM. For large lattices, $C_L$ for
the two-dimensional systems is practically identical to its ladder value. 
This suggests
that the correlation length is very short. In this regime  short-ranged
plaquettes are dominant in the exact wave-function, hence 
${\tilde \Psi}$ is an excellent variational approximation. 

Recently a variational wave function based on doping an orthogonal dimer wave 
function has been used to explore the nature of an eventual superconductive 
state in $SrCu_2(BO_3)_2$ \cite{liu}. The estimated value of the couplings 
in this compound are $J_1\approx 85 K$ and $J_2=54 K$ \cite{ueda}. This 
places it at the boundary of the orthogonal dimer phase, not very far for 
the plaquette phase. Given the possible incertitude in this estimation and
the fact that these values may be affected by doping, it is worth exploring
superconductivity upon doping the plaquette ground state as well.

To summarize, I have argued that the DMRG technique is a natural
approach to study spin-Peierls phases that spontaneously arise in
frustrated quantum antiferromagnets. 
I have used the two-step DMRG to confirm the nature
of the controversial phase between the N\'eel and dimer
phases. The phase diagram of the SSM bears some similarity to that
of the checkerboard model \cite{checkerboard}. In the checkerboard
model, an additional N\'eel phase was recently found between the
plaquette and the crossed dimer phases \cite{sfb,checkerboard}.
It is quite possible that this additional phase exists in the
SSM in the vicinity of $J_2 \approx 1.5 J_1$. Unfortunately, in the
SSM, unlike the checkerboard model, the starting ladder does not have
the full symmetry of the bond pattern of the 2D lattice. As the dimer
phase is approached, the variational wave function used in this study is not 
optimal, as it leaves half of the spins unpaired in the orthogonal dimer
phase. It will be more advantageous to start with the orthogonal
dimer pattern shown in Fig.\ref{lattice}. 

\begin{acknowledgments}
 This work was supported by the NSF Grant No. DMR-0426775. I am grateful 
to Prof. V. Lieberman for reading the manuscript.
\end{acknowledgments}

\end{document}